\documentclass[12pt]{iopart}
\usepackage{cite,./mcite}
\usepackage{array}
\usepackage{epsfig}
\def\lsim{\mathrel{\rlap{\lower4pt\hbox{\hskip1pt$\sim$}}
    \raise2pt\hbox{$<$}}} 
\def\gsim{\mathrel{\rlap{\lower4pt\hbox{\hskip1pt$\sim$}}
    \raise2pt\hbox{$>$}}} 
\newcommand{\abst}{\mbox{$|t|$}}

\newcommand{\pom}{{\rm I\! P}}

\newcommand{\mev}{{\rm Me}\kern-1.pt{\rm V}}
\newcommand{\gev}{{\rm Ge}\kern-1.pt{\rm V}}
\newcommand{\gevsq}{\mbox{$\mathrm{{\rm Ge}\kern-1.pt{\rm V}}^2$}}

\newcommand{\rhoz}{\mbox{$\rho^0$}}
\newcommand{\jpsi}{\mbox{$J/\psi$}}

\newcommand{\qsq}{\mbox{$Q^2$}}

\begin{document}

\title{Experimental Results on the Diffractive Production of Light Vector Mesons}

\author{James A.~Crittenden
\footnote{On leave from Phys. Inst. Bonn, Nu{\ss}allee 12, 53115 Bonn}
\\
(for the H1 and ZEUS collaborations)}

\address{Deutsches Elektronen-Synchrotron, Notkestra{\ss}e 85, 22603 Hamburg}

\begin{abstract}
We discuss results on the diffractive production of the vector mesons {\rhoz}, $\phi$ and $\omega$ reported by the H1 and ZEUS collaborations at HERA. Unique to 
such studies is the experimental accessibility to the polarization of the
vector mesons and hence to the spin-density matrix elements arising
in vacuum-exchange processes. We emphasize the relation between the observed
dependence on momentum transfer and the 
polarization state of the vector meson.
The diffractive nature of the production mechanism is investigated via 
extraction of the Pomeron trajectory at high {\qsq}. Flavor symmetry
is observed in the $\phi/{\rhoz}$ ratios in the same region of momentum
transfer where the power-law scaling becomes similar. The multivariable helicity analyses impose stringent constraints on models for the vacuum-exchange production mechanism. Semi-exclusive photoproduction of transverse
{\rhoz} and {$\phi$} mesons at momentum transfers 
far exceeding their mass scale 
exhibit a hard scaling behavior which appears to violate the 
QCD helicity selection rules in a two-gluon exchange model. 
\\*[10mm]
Contributed to the proceedings of  the Ringberg Workshop on New
Trends in HERA Physics 2001, 17--22 June 2001, Ringberg Castle,
Tegernsee, Germany.
\end{abstract}



\maketitle

\section{Introduction}
Investigations of the exclusive production of vector  mesons 
in electron-proton interactions
by the H1 and ZEUS collaborations at HERA have yielded
detailed information on vacuum-exchange dynamics, since the high
flux of real and virtual photons from the electron beam provides a means
of investigating such processes with high precision. 
The particular interest in the light
vector mesons {\rhoz}, $\phi$ and $\omega$ lies in the opportunity to
reach kinematic regions of momentum transfer in which effects of
binding are small and the fundamental scaling properties of the underlying
vacuum-exchange process are revealed. In contrast to
studies of inclusive processes, decay-angle analyses in exclusive
and semi-exclusive processes yield detailed information on the 
spin-density matrix, i.e. the helicity-transfer characteristics of
the production mechanism. Since these are related on very general theoretical
grounds to the scaling properties in field theoretical 
descriptions~\cite{pr_22_2157,prep_112_173}, stringent constraints
on the phenomenological interpretation of strong vacuum-exchange 
mechanisms are obtained. In recent years, much theoretical work has
been done on such interpretations, both for vector-meson electroproduction
at high photon virtuality~\cite{pr_50_3134,pr_56_2982},
and for photoproduction at high
momentum transfer~\cite{pr_53_3564,pr_54_5523,pl_478_101}.

This report concerns recent results on elastic {\rhoz} electroproduction
from longitudinal and transverse virtual photons, including a
Regge analysis at high {\qsq}, investigations of the dependence of
the spin-density matrix elements on proton dissociation and the momentum
transferred to the proton, as well as scaling and helicity-transfer properties
of {\rhoz} and $\phi$ photoproduction at high momentum transfer.
The results address quantitatively a variety of theoretical concerns. 
By comparing the properties of the photon/vector-meson transition for elastic
electroproduction to those for the 
proton-dissociative process, one tests the interpretation
of the process as diffraction, since any difference implies a correlation 
between the photon/vector-meson and the hadronic vertices.
Any {\qsq} dependence
in an extracted Regge trajectory challenges a canonical assumption of 
fixed poles in Regge theory. The dependence of the forward 
cross sections on {\qsq} confront recent QCD descriptions of diffractive processes
which predict asymptotic dominance of the production of longitudinal vector mesons
from longitudinal virtual photons, with a Q$^{-6}$ dependence modified by the
{\qsq} dependence in the gluon density and in the strong coupling. The transverse
cross section is expected to be suppressed by a factor of {\qsq}, though
corrections due to the endpoints 
of the quark momentum distributions in the meson mitigate this expectation. 
A related line of argument has been proposed to predict the dominance of
longitudinal vector mesons at high {\abst} in diffractive photoproduction,
where $t$ is the square of the momentum transferred to the proton. Again,
production of transverse vector mesons is expected to be suppressed by a factor
of $t$, resulting in a prediction of ${\rmd \sigma}/{\rmd  t} \propto {(-t)}^{-4}$, 
modified by the $t$ dependence in the gluon density and in the 
strong coupling. These helicity selection rules in the dependence on momentum
transfer are straightforward consequences of the two-gluon model for diffractive 
processes~\cite{diehlringberg}.

\section{Scaling laws and helicity selection in vector-meson electroproduction}

\begin{figure}[htbp]
\begin{minipage}{\textwidth}
\begin{center}
  \epsfig{file=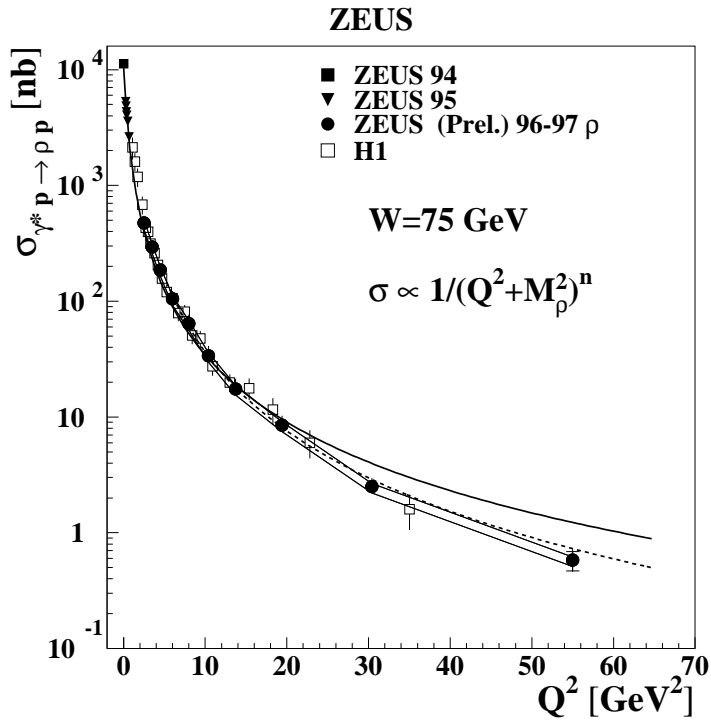,width=0.6\textwidth}
\end{center}
\end{minipage}
  \caption{
Elastic {\rhoz} electroproduction cross section as a function of {\qsq}. 
The lines show the result of fits to the form $(\qsq + {\rm M}^2_{\rho})^{-n}$.
The solid line indicates the result of the fit to all data points, while
the dashed line is the result restricted to the points at $\qsq \geq 5~\gevsq$.
Also indicated is an 
additional normalization uncertainty associated with the 
subtraction of the proton-dissociative background.}
  \label{fig:dsigdq2}
\end{figure}

\begin{figure}[htbp]
\begin{minipage}{\textwidth}
\begin{center}
  \epsfig{file=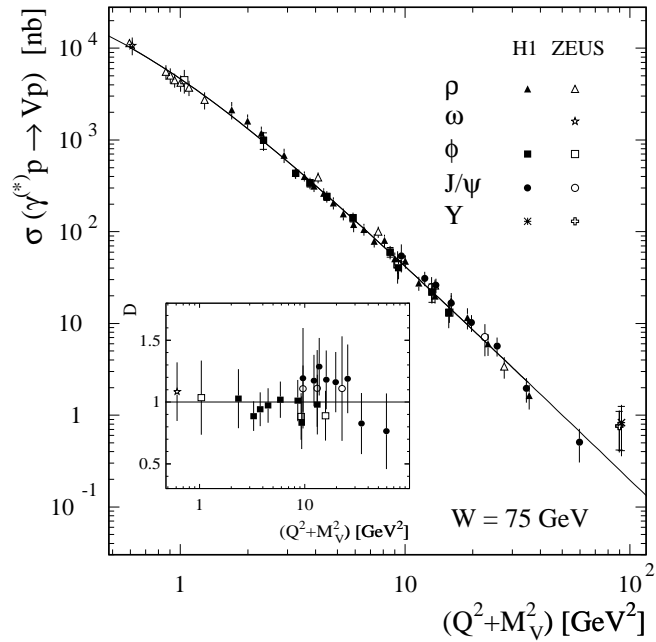,width=0.6\textwidth}
\end{center}
\end{minipage}
  \caption{
Elastic {\rhoz} electroproduction cross section as a function of $(\qsq + {\rm M}^2_{\rm V})$ for ${\rm V}={\rhoz}, \phi, \omega, \jpsi$ and $\Upsilon$, scaled
according to the proportions 9:2:1:8:2, which correspond to the quark-charge
content of the vector mesons. The error bars indicate the quadratic sum of
statistical and systematic uncertainties. The curve indicates the result of
a fit to the ZEUS and H1 {\rhoz} data. The ratio $D$ of the
$\phi$, $\omega$ and {\jpsi} cross sections to the fit 
result is shown in the inset.
}
  \label{fig:vmscale}
\end{figure}

\begin{figure}[htbp]
\begin{minipage}{\textwidth}
\begin{center}
  \epsfig{file=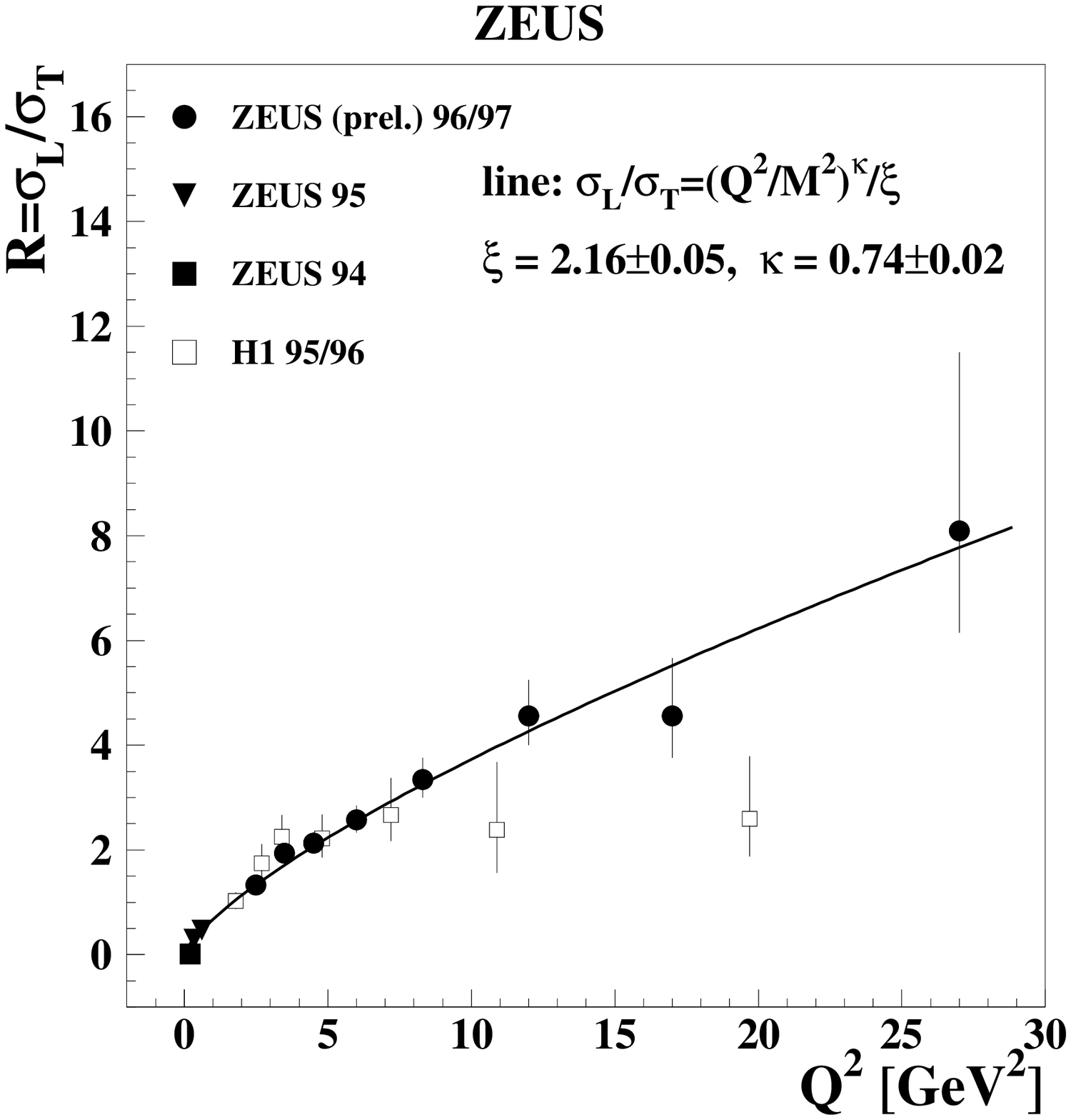,width=0.6\textwidth}
\end{center}
\end{minipage}
  \caption{
Ratio of the longitudinal and transverse elastic {\rhoz} electroproduction
cross sections, \mbox{$R=\sigma_{\rm L}/\sigma_{\rm T}$},
as a function of {\qsq}. The solid line shows the result of a fit to the
form $R=(Q^2/M^2_{\rho})^\kappa{/\xi}$.
}
  \label{fig:rvsq2}
\end{figure}

\begin{figure}[htbp]
\begin{minipage}{\textwidth}
\begin{center}
  \epsfig{file=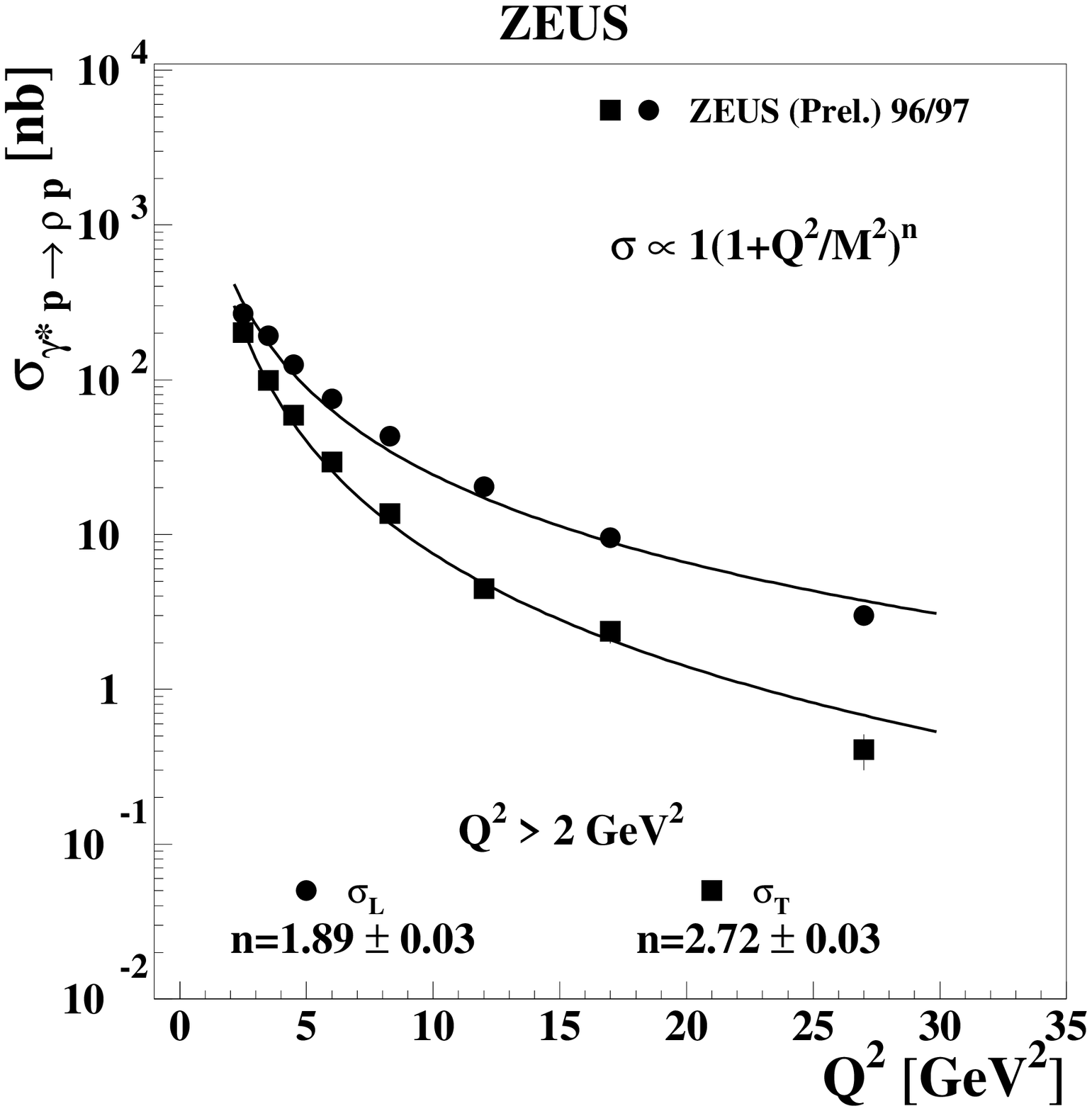,width=0.6\textwidth}
\end{center}
\end{minipage}
  \caption{
The longitudinal and transverse elastic {\rhoz} electroproduction
cross sections
as a function of {\qsq}. The solid lines shows the result of fits to the
form $(\qsq + {\rm M}^2_{\rho})^{-n}$.
}
  \label{fig:qsigLT}
\end{figure}

Figure~\ref{fig:dsigdq2} 
shows the recently obtained 
{\qsq} dependence of the total {\rhoz} elastic cross section
presented by the ZEUS collaboration~\cite{eps01_594}.
A fit to the form $(\qsq + {\rm M}^2_{\rho})^{-n}$ shows that the power $n$
varies with the lower bound on {\qsq} chosen for the fit. For 
lower bounds greater than $5~\gevsq$, 
the fit is statistically consistent with a power of $2.36 \pm 0.04$.
The H1 collaboration has remarked upon the scaling of the total cross section,
pointing out that the $\phi$ and {\jpsi} cross sections exhibit 
scaling behavior remarkably similar to that for the {\rhoz} 
when weighted with normalization
factors corresponding to the quark-charge content of the meson, as shown in 
Fig.~\ref{fig:vmscale}~\cite{pl_483_360}. 
However, each of the above 
observations is difficult to interpret phenomenologically, since
the contributions from longitudinal and transverse photons are expected
to scale differently, and the relative contributions are known to vary
rapidly with {\qsq}, as shown in Fig.~\ref{fig:rvsq2}~\cite{eps01_594}.
This measurement of the ratio \mbox{$R=\sigma_{\rm L}/\sigma_{\rm T}$}
allowed extraction of the scaling properties of 
$\sigma_{\rm L}$ and $\sigma_{\rm T}$
separately,
as shown in Fig.~\ref{fig:qsigLT}~\cite{ichep00_439}.
The separation of the scaling properties of the cross sections
for longitudinal and transverse photons was obtained by analysis
of the polar decay-angle distribution.
The hard scaling behavior is impressive, since the longitudinal cross
section is even harder than a point-like-scattering (Rutherford) 
cross section. In the context of
QCD calculations which model vacuum exchange as the exchange
of a color-neutral pair of gluons, 
this indicates that the {\qsq} dependence of the
square of the gluon density compensates the leading $Q^{-6}$ behavior
by more than a factor of {\qsq}. Also remarkable is the hard behavior
of the transverse cross section, which encourages a leading-order field
theoretical interpretation as well. In the case of the production of 
transverse vector mesons, however, the QCD calculation is complicated
by contributions arising from the endpoints of the quark momentum distribution
in the vector meson~\cite{pr_56_2982},
resulting in the necessity for modelling 
non-perturbative effects~\cite{pr_55_4329,*pr_58_114026}. 
It has been experimentally established that the spin-flip
contributions are rather small ($\simeq$~10\%)~\cite{epj_12_393,*epj_10_373},
so the observed {\qsq} dependences in the cross sections
for longitudinal and transverse photons are each determined by
the helicity-conserving amplitudes to a good approximation.

\begin{figure}[htbp]
\begin{minipage}{\textwidth}
\begin{center}
  \epsfig{file=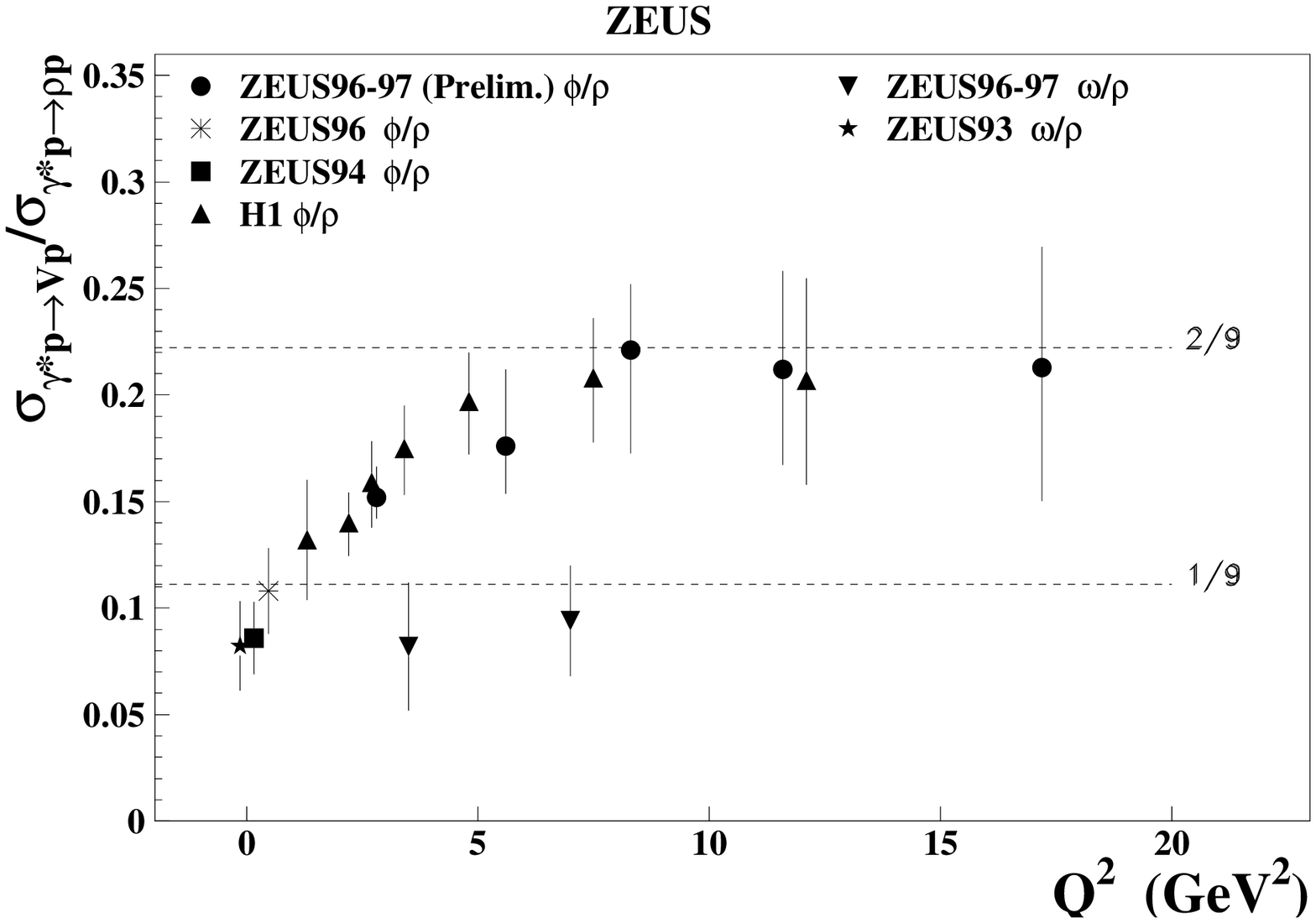,width=0.6\textwidth}
\end{center}
\end{minipage}
  \caption{
Ratios of the $\phi$ and $\omega$ elastic electroproduction cross sections
to that for the {\rhoz} as a function of {\qsq}. The lines indicate the
values for the ratios derived from the quark-charge content of the 
vector mesons.}
  \label{fig:ratiovsq2}
\end{figure}

The ratio of $\phi$ to {\rhoz} elastic electroproduction is
shown in Fig.~\ref{fig:ratiovsq2}~\cite{dis01_kreisel}. 
Of particular
interest is the observation that the {\qsq} dependence for these
two vector mesons becomes similar at the same scale at which the
cross section ratio reaches the value obtained from simple quark-charge 
counting.
Since the ratio $R$ for the $\phi$ is known to exhibit a {\qsq}
dependence similar to that 
for the {\rhoz}~\cite{pl_483_360,*ichep98_793}, we can conclude that
$\sigma_{\rm L}$ and $\sigma_{\rm T}$ show a hard scaling behavior for the $\phi$ as well,
and that the observed scaling is a property of the vacuum-exchange process
little modified by meson binding effects. 

\newpage
Figure~\ref{fig:ratiovsq2} also shows that the $\omega$/{\rhoz} ratio, which
reached the value derived from quark-charge counting at low {\qsq}, does
not rise with {\qsq} in this range. One can conclude either that some
$\phi$ suppression mechanism is active at low {\qsq}, $or$ that there
is an additional 
production mechanism which is similar for the {\rhoz} and $\omega$ at low {\qsq} 
but absent in the case of the $\phi$.

\begin{figure}[htbp]
\begin{minipage}{\textwidth}
\begin{center}
  \epsfig{file=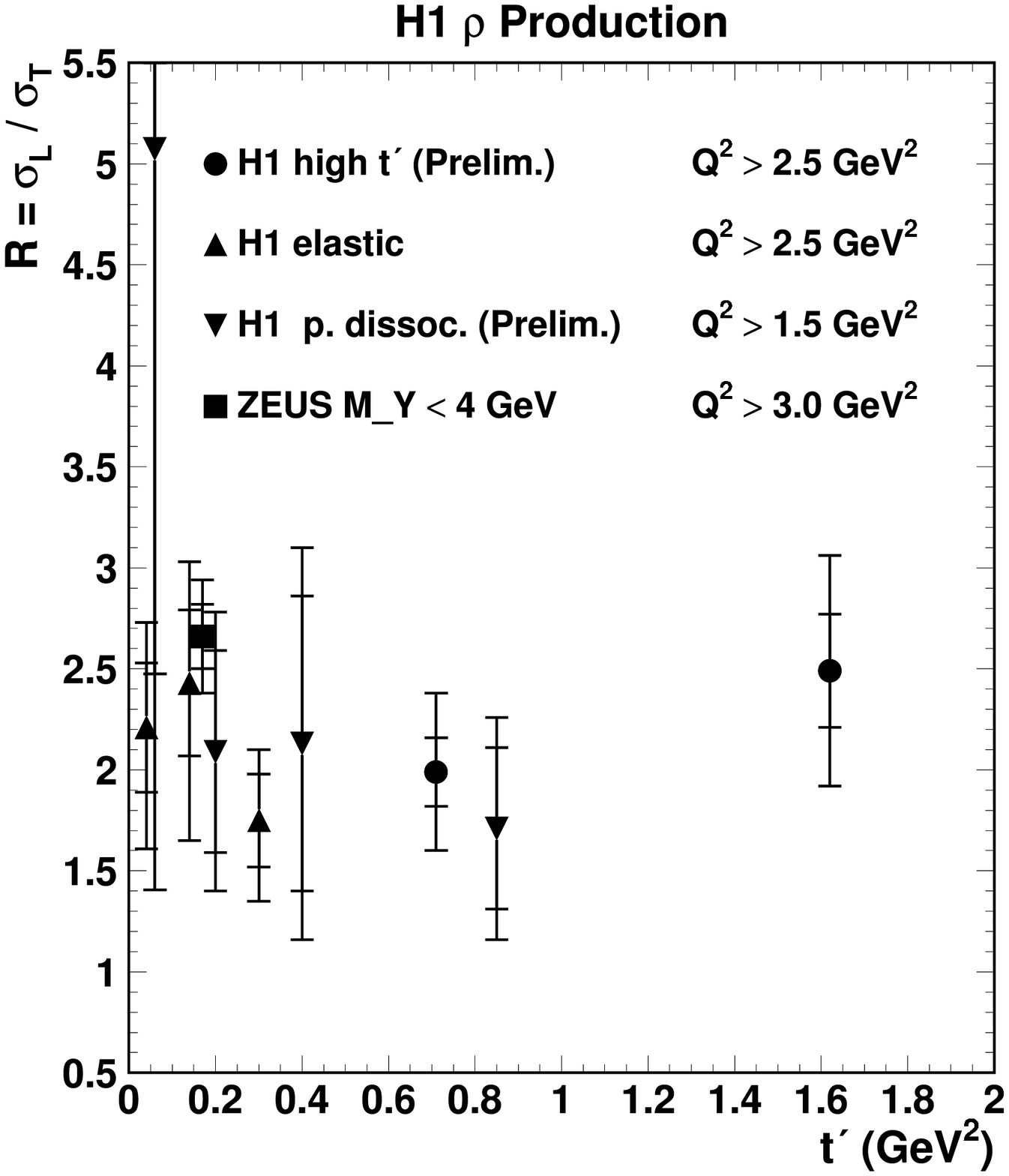,width=0.43\textwidth}
  \epsfig{file=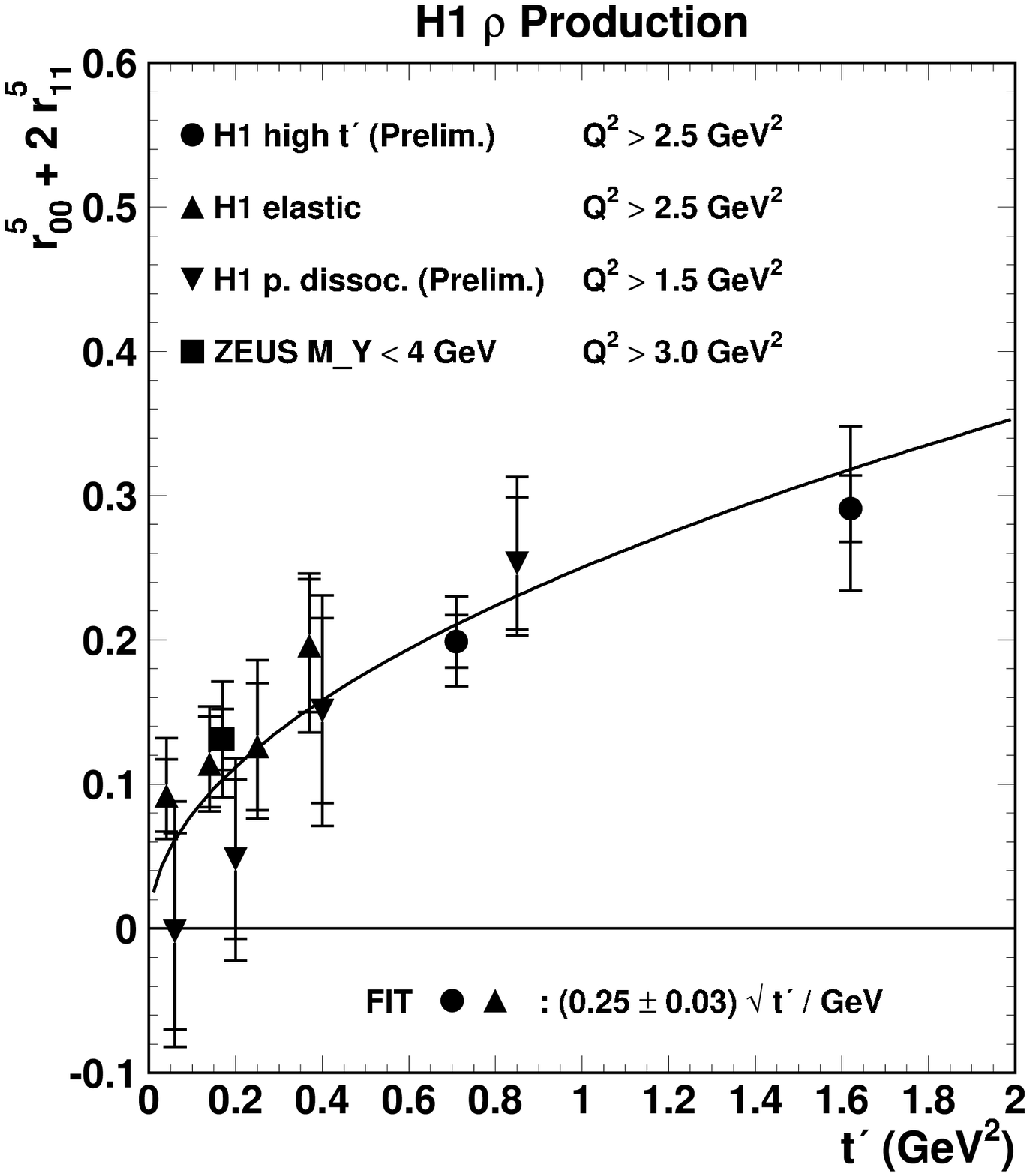,width=0.43\textwidth}
\end{center}
\end{minipage}
  \caption{
Ratio of the longitudinal and transverse {\rhoz} electroproduction
cross sections, \mbox{$R=\sigma_{\rm L}/\sigma_{\rm T}$},
and the combination of spin-density matrix elements
$r^{5}_{00}+r^{5}_{11}$
as a function of $t^\prime$. 
}
  \label{fig:h1hight}
\end{figure}

The H1 collaboration  has investigated the dependence of the ratio
$R$ and the combination of spin-density matrix elements
$r^{5}_{00}+r^{5}_{11}$ on the momentum transfer to the
proton~\cite{eps01_796}. Figure~\ref{fig:h1hight} 
shows the dependence on $t^\prime$, which denotes 
the momentum transferred to the proton after subtraction of
the kinematic lower bound. 
The
ratio $R$ shows no strong dependence on $t^\prime$ for {\qsq} greater than
a few {\gevsq}. 
The sum
$r^{5}_{00}+r^{5}_{11}$ must vanish for \mbox{$s$-channel}
amplitudes which conserve helicity~\cite{np_61_381}
in the photon/vector-meson transition. 
Such conservation of helicity 
is a simple consequence of angular momentum conservation at zero momentum transfer.
These results show evidence for a nonzero contribution from
helicity-violating amplitudes which increases with $t^\prime$.

\section{The Pomeron trajectory measured in {\rhoz} electroproduction}
The ZEUS collaboration has extracted a measurement of the Pomeron
trajectory in {\rhoz} electroproduction at {\qsq} values of 
3.5 and 10.0~{\gevsq} by measuring the energy dependence of the
elastic cross section in four bins of $t$ up to a value
of 0.6~\gevsq\cite{eps01_594}.
They find a value for the intercept of
\mbox{$\alpha_{\pom}(t=0)=1.14\pm0.01{\rm (stat)}\pm0.03{\rm (sys)}$}
and a slope of 
\mbox{$\alpha^\prime_{\pom}=0.04\pm0.07{\rm (stat)}^{+0.13}_{-0.04}{\rm (sys)}$}.
One can conclude that the trajectory for this process has a higher intercept
and less shrinkage than the soft Pomeron trajectory which successfully
describes hadronic cross sections at low momentum transfer.

\section{Vector-meson photoproduction at high momentum transfer}
\begin{figure}[htbp]
\begin{minipage}{\textwidth}
\begin{center}
  \epsfig{file=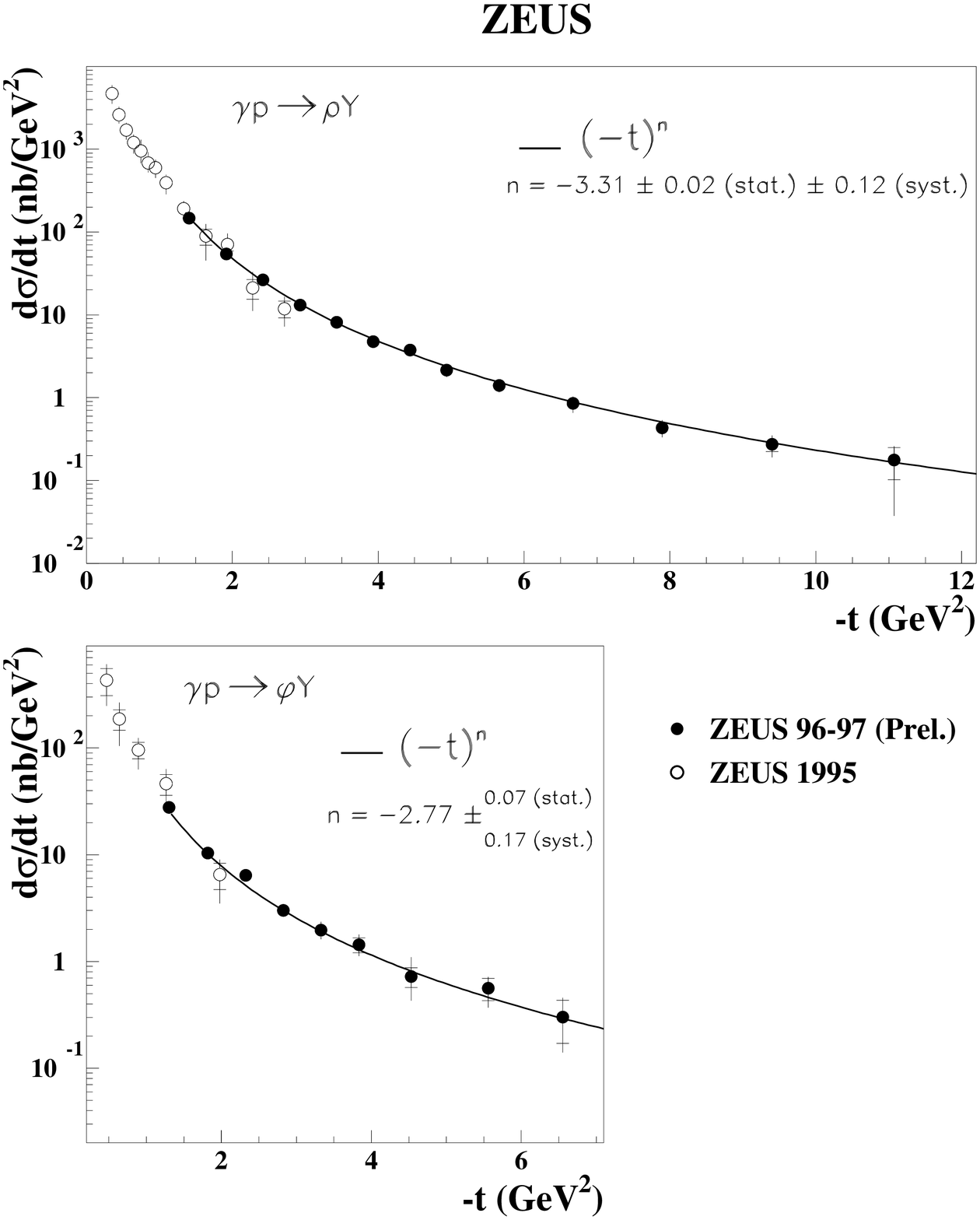,width=0.6\textwidth}
\end{center}
\end{minipage}
  \caption{
Differential cross sections ${\rmd \sigma}/{\rmd  t}$
for proton-dissociative photoproduction of {\rhoz} and $\phi$ mesons.
The open circles show the results of Ref.~\protect\cite{epj_14_213}.
The lines show the results of fits of the form $\rmd {\sigma}/\rmd t \propto (-t)^{-n}$ to the solid
circles.
}
\label{fig:dsigdt}
\end{figure}

The ZEUS collaboration has extended its previous measurement of
proton-dissociative
vector-meson photoproduction~\cite{epj_14_213} 
to higher momentum transfer (${\abst}\lsim 11~{\gevsq}$)
by exploiting
the higher statistics of the data set recorded in 1996/1997~\cite{eps01_556}. 
A small
electron calorimeter positioned near the electron beam-pipe 44~m from
the interaction point served to select events initiated by
quasi-real photons in the process 
\mbox{$\gamma + p \rightarrow {\rm VM}+ Y$,} where $Y$ represents a dissociated 
state of the proton. The resulting range in $\gamma p$ center-of-mass energy
extends from 80 to 120~\gev. Since the transverse
momentum of the final-state positron was required to be
small 
by the geometrical acceptance of the calorimeter (${\qsq}<0.02~{\gevsq}$), 
the transverse momentum of the vector meson, $p_{\rm t}$, provided an accurate estimate of the square of the momentum
transferred to the proton via $t\simeq -p^2_{\rm t}$. 
The differential cross sections
${\rmd \sigma}/{\rmd  t}$ were
obtained in the region ${\abst}>1.2~\gevsq$ and are shown in Fig.~\ref{fig:dsigdt}. 
The measurements provide good sensitivity
to the observed power-law dependence 
${\rmd \sigma}/{\rmd  t} \propto {(-t)}^{-n}$~\cite{jacaps}.
Over the region in momentum transfer covered by the data, the power is 
found to be \mbox{$n\,=\,3.31\pm 0.02{\rm (stat)}\pm 0.12{\rm (sys)}$} for the {\rhoz} and \mbox{$n\,=\,2.77\pm 0.07{\rm (stat)}\pm 0.17{\rm (sys)}$} for the
$\phi$~\cite{eps01_556}. 

\begin{figure}[htbp]
\begin{minipage}{\textwidth}
\begin{center}
  \epsfig{file=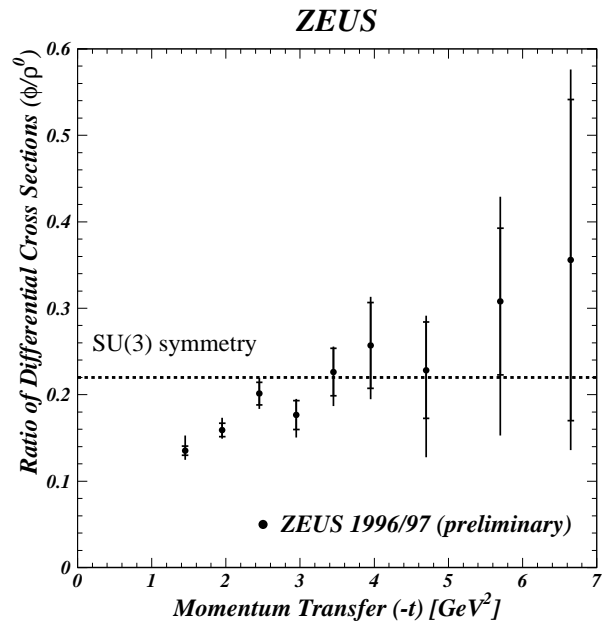,width=0.55\textwidth}
\end{center}
\end{minipage}
  \caption{
Ratio of the differential cross section ${\rmd \sigma}/{\rmd  t}$
for proton-dissociative $\phi$ photoproduction
to that for the {\rhoz} as a function of $t$. The line indicates the
value for the ratio derived from the quark-charge content of the 
vector mesons.
}
  \label{fig:phirhovst}
\end{figure}

Figure~\ref{fig:phirhovst} 
shows the ratio of {\rhoz} and {$\phi$} cross sections
as a function of $t$. Just as in the case of electroproduction, one
observes that the
momentum-transfer dependence for these
two vector mesons becomes similar at the same scale at which the
cross section ratio reaches the value obtained from quark-charge 
counting. Such an observation yields valuable information concerning
the nature of the short-distance coupling which is  
free of the large uncertainties arising from meson-binding effects
in calculations of the absolute 
magnitudes of the cross sections.

\begin{figure}[htbp]
\begin{minipage}{\textwidth}
\begin{center}
  \epsfig{file=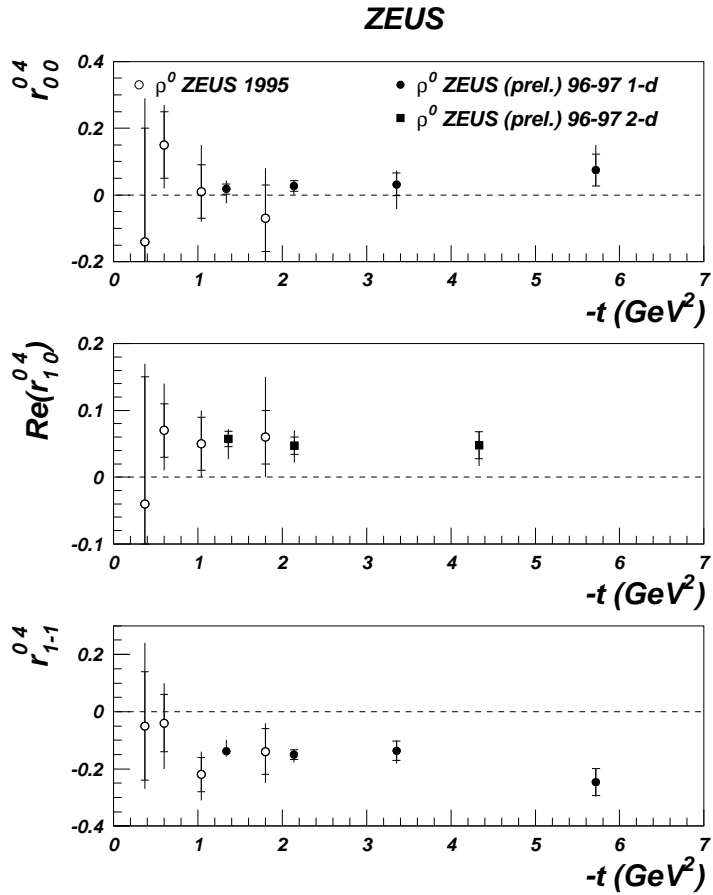,width=0.6\textwidth}
\end{center}
\end{minipage}
  \caption{
The values for the combinations of spin-density matrix elements
$r^{04}_{00}$, \mbox{Re $r^{04}_{10}$} and $r^{04}_{1-1}$ for
proton-dissociative {\rhoz} photoproduction derived from fits
to decay-angle distributions, plotted as functions of $t$.
The open circles indicate the measurements of Ref.~\protect\cite{epj_14_213}.
}
  \label{fig:rhoelements}
\end{figure}

Figure~\ref{fig:rhoelements} shows the
values for the combinations of spin-density matrix elements measured in 
the $s$-channel helicity frame
$r^{04}_{00}$, \mbox{Re $r^{04}_{10}$} and $r^{04}_{1-1}$ as a function of
$t$ obtained from this data sample. 
The helicity-conserving 
(transverse-to-transverse) process clearly dominates,
as shown by the small value of $r^{04}_{00}$, which is the probability
for longitudinal production, and by the small value of \mbox{Re $r^{04}_{10}$},
which is sensitive to the interference between the helicity-conserving
(\mbox{$T\rightarrow T$}) and single-flip (\mbox{$T\rightarrow L$}) amplitudes.
The statistical power of the data
also allow accurate determination of a non-zero double-flip contribution,
as shown by the non-zero value of $r^{04}_{1-1}$ at high {\abst}, since it
arises from the interference between the non-flip and double-flip amplitudes.

It should be noted that these results represent the 
first measurements of light-vector-meson photoproduction 
at values of $t$ comparable to those of {\qsq} 
in inclusive deep inelastic electron-proton
scattering which revealed the existence of
fractionally charged proton constituents at SLAC in 1967. 
In contrast to the deep inelastic scattering process, this vacuum-exchange 
reaction
is presumably sensitive to strong couplings rather than to
the electric charges of the proton constituents. A transition to power-law
scaling is observed, and the momentum-transfer dependence for the {\rhoz} 
becomes similar to that for the {$\phi$} at the same scale at which the
cross section ratio reaches the value obtained from quark-charge 
counting. Since this is also the region in which the momentum transfer exceeds
the masses, one can plausibly assume that the observed dependence
characterizes the underlying dynamics independent of meson-binding effects.
Given the phenomenological successes of QCD descriptions of diffractive
electroproduction of longitudinal vector mesons 
during the past few years, it is remarkable that
the observed power for the dominant transverse-to-transverse amplitude
in photoproduction
is lower than expected from two-gluon-exchange
calculations~\cite{pl_478_101}. These calculations
yield a leading $(-t)^{-4}$ dependence, further
steepened by the running of the strong coupling constant: 
${\rmd \sigma}/{\rmd  t} \propto \alpha^4_{\rm s}/{(-t)}^{4}$.
It is interesting to note that this expectation for the leading
power behavior is not specific to QCD. Chernyak and Zhitnitsky
have
pointed out, for example, that these rules hold for the exchange of any 
vector or axial-vector current or for any Lagrangian 
with dimensionless coupling constants~\cite{prep_112_173}.

\section{Conclusions}
Investigations of the diffractive production of light vector mesons at HERA
provide detailed information on the dynamics
which govern vacuum-exchange processes. The dependence on proton-dissociation 
of the helicity-transfer properties of the virtual-photon/vector-meson
transition has been shown to be weak, supporting
the characterization of vector-meson electroproduction in this
kinematic region as a diffractive process. A
Pomeron trajectory has been measured in exclusive {\rhoz} electroproduction
at high photon virtuality for the first time. This trajectory exhibits
a higher intercept and weaker slope than does the trajectory measured 
in {\rhoz} photoproduction, which was similar to that
determined in soft hadronic processes.

High-statistics measurements at values of both {\qsq} and $t$ far exceeding 
the light-meson mass scale have been obtained. The expectation that
meson-binding effects are small in this region are supported by the
observation that  the momentum-transfer dependence for the {\rhoz} 
becomes similar to that for the {$\phi$} at the same scale at which the
cross section ratio reaches the value obtained from quark-charge 
counting. The common power behavior may thus be interpreted as characterizing
the vacuum-exchange dynamics alone.

Decay-angle analyses have 
experimentally distinguished the momentum-transfer scaling properties
of the various helicity amplitudes.
The {\qsq} dependence for the longitudinal elastic {\rhoz} electroproduction
has been found to follow a power law with a power approximately one unit
lower than observed for the transverse cross section, 
as expected on very general theoretical
grounds. However, results on the semi-exclusive
photoproduction of {\rhoz} and {$\phi$} mesons at high momentum transfer
show a remarkably hard power law to govern the short-distance
transverse-to-transverse vacuum-exchange amplitude. The fundamental
nature of the theoretical basis which leads to the expectation of a softer
spectrum is likely to motivate an intensification of the 
already widespread phenomenological activities which address the 
many experimental results on the diffractive production of vector mesons
at HERA.

\section{Acknowledgments}
This work is supported by the Federal Ministry for Education and Research of Germany.
%
\section*{References}
\bibliographystyle{./bib/thera}
{\raggedright
\bibliography{./bib/theratdr,./bib/crittenden,./bib/eps01,./bib/dis01}
}
\end{document}